%% file: main.tex
\documentclass[sigconf,authorversion,nonacm]{acmart}
\AtBeginDocument{%
  }

\acmConference[CHI '27]{ACM CHI Conference on Human Factors in Computing Systems}{April 2027}{}
\acmDOI{}
\acmISBN{}

\usepackage{multirow}
\usepackage{color}
\usepackage{colortbl}
\usepackage{stfloats}
\usepackage{subcaption}
\usepackage{enumitem}
\usepackage{graphicx}
\usepackage{pifont}
\usepackage{tabularx}
\usepackage{soul}
\usepackage{tikz}
\usepackage{etoolbox}
\usepackage{kotex}
\usepackage{tipa}
\usepackage[table]{xcolor}
\usepackage{booktabs}
\usepackage{siunitx}
\usepackage{accents}
\usepackage{svg}
\usepackage{longfbox}
\usepackage{tcolorbox}
\usepackage{array}

\definecolor{parentcolor}{HTML}{cfe5cc}
\definecolor{childcolor}{HTML}{edd8b9}
\definecolor{expertcolor}{HTML}{e5ccef}

\newcommand{\participantbox}[2]{%
  \tcbox[
    on line,
    boxrule=0pt,
    arc=2pt,
    boxsep=0pt,
    left=2pt,
    right=2pt,
    top=1.5pt,
    bottom=1.5pt,
    colback=#1,
    colframe=#1
  ]{#2}%
}

\newcommand{\notherapy}[1]{\participantbox{parentcolor}{P#1}}
\newcommand{\notherapykid}[1]{\participantbox{parentcolor}{C#1}}

\newcommand{\yestherapy}[1]{\participantbox{childcolor}{P#1}}
\newcommand{\yestherapykid}[1]{\participantbox{childcolor}{C#1}}

\newcommand{\formative}[1]{\participantbox{expertcolor}{F#1}}
\newcommand{\expert}[1]{\participantbox{expertcolor}{E#1}}

\begin{document}

\title{MORA: Designing an AI-Mediated Role-Playing System for Child--Caregiver Speech Sound Practice}

\author{Sumin Hong}
\email{shong6@nd.edu}
\affiliation{%
  \institution{University of Notre Dame}
  \city{Notre Dame}
  \state{Indiana}
  \country{USA}
}

\author{Yinuo Yang}
\email{yyang39@nd.edu}
\affiliation{
  \institution{University of Notre Dame}
  \city{Notre Dame}
  \state{Indiana}
  \country{USA}
}

\author{Jiawen Li}
\email{jliu46@nd.edu}
\affiliation{
  \institution{University of Notre Dame}
  \city{Notre Dame}
  \state{Indiana}
  \country{USA}
}

\author{Xavier Briggs}
\email{xbriggs@nd.edu}
\affiliation{
  \institution{University of Notre Dame}
  \city{Notre Dame}
  \state{Indiana}
  \country{USA}
}

\author{Qingxiao Zheng}
\email{qingxiao@buffalo.edu}
\affiliation{%
  \institution{University at Buffalo, SUNY}
  \city{Buffalo}
  \state{New York}
  \country{USA}
}

\author{Rafiah Ansari}
\email{rafiah.ansari.2@citystgeorges.ac.uk}
\affiliation{%
  \institution{City, University of London}
  \city{London}
  \country{United Kingdom}
}

\author{JinJun Xiong}
\email{jinjun@buffalo.edu}
\affiliation{%
  \institution{University at Buffalo, SUNY}
  \city{Buffalo}
  \state{New York}
  \country{USA}}
  
\author{Yao Du}
\email{yaodu@usc.edu}
\affiliation{%
 \institution{University of Southern California}
 \city{Los Angeles}
 \state{California}
 \country{USA}}

\author{Toby Jia-Jun Li}
\email{toby.j.li@nd.edu}
\affiliation{
  \institution{University of Notre Dame}
  \city{Notre Dame}
  \state{Indiana}
  \country{USA}
}
\renewcommand{\shortauthors}{Hong et al.}
\DeclareRobustCommand{\sysname}{\textbf{MORA}}
\newcommand{\etal}{\textit{et al.}}
\renewcommand{\arraystretch}{1.2}

\begin{abstract}

Treating speech sound disorder (SSD) requires repeated practice across linguistic complexity levels. However, in the home setting, caregivers lack the expertise to judge children's productions to sustain intensive practice.
Using a formative study with six speech-language pathologists (SLPs) and an expert walkthrough with seven SLPs, we present \sysname{}, an AI-mediated role-playing story-based system in which target-bearing words recur as characters, locations, and props, eliciting productions under two response constraints while retaining recordings for clinician review.
Because ASR artifact, more than child error, dominated phoneme-distance scores of disordered recordings at low error levels, the dashboard abstains rather than classifies there. 
A lab-based user study with 10 child–caregiver dyads (five reporting SSD or therapy history) surfaced three tensions: connected speech and comprehensible tasks, a responsive signal and the authority it accrues, and child autonomy and the coordination it requires. 
We discuss design implications for revisable configuration, scope-bounded feedback, and coordination support.
\end{abstract}



\begin{teaserfigure}
  \includegraphics[width=\textwidth]{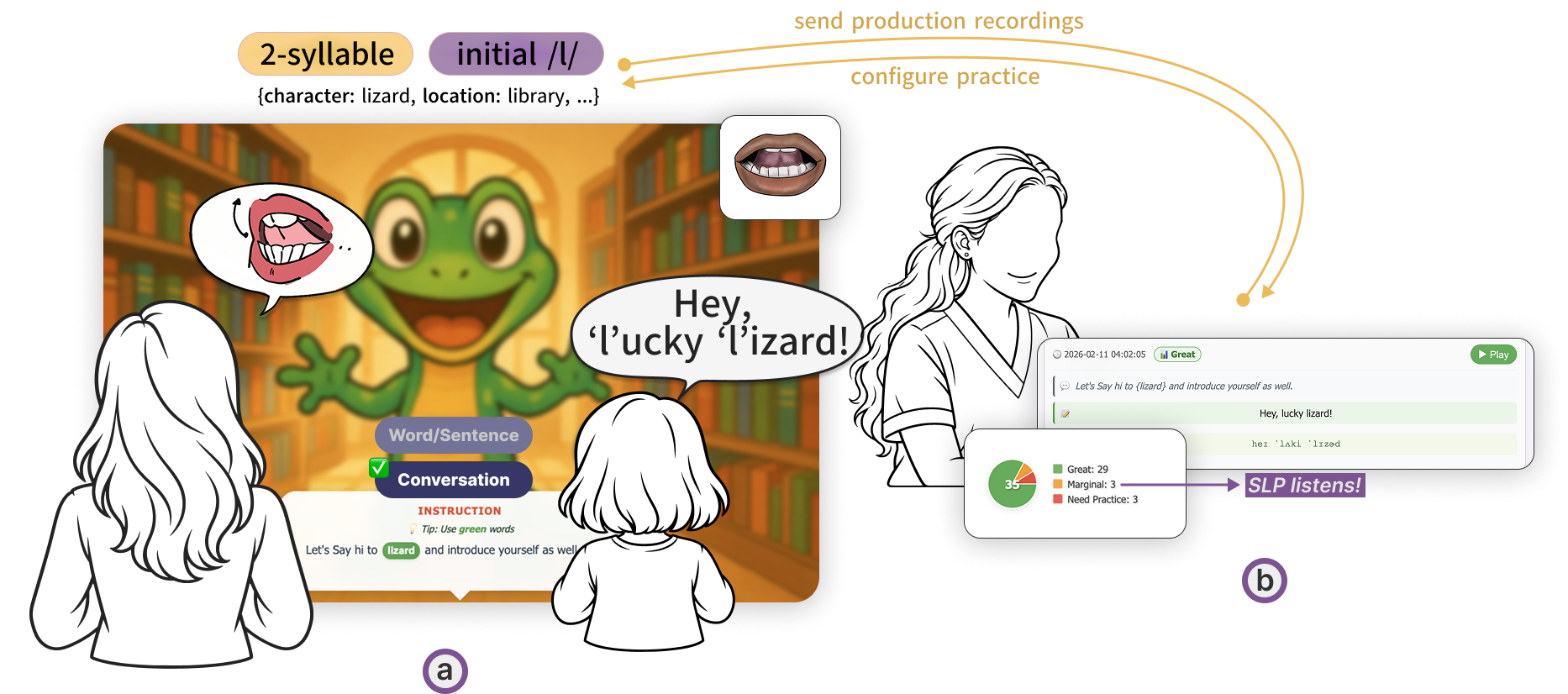}
  \caption{\sysname{} supports children's speech-sound practice as an interactive, role-playing story. Configured targets (here, 2-syllable words with initial /l/) become recurring story roles, and the child practices with caregiver support and articulatory cues, responding in Word/Sentence or Conversational Mode (a). Each recording is retained with its prompt, transcripts, and SLP-calibrated label on a clinician-facing dashboard, where \textit{Marginal} recordings are routed to the clinician for listening (b). The arrows depict the intended workflow: recordings flow to professional review, and review informs the next practice configuration.}
\end{teaserfigure}


\maketitle

\input{sections/1-intro}
\input{sections/2-related-work}
\input{sections/3-design}
\input{sections/4-system}
\input{sections/5-calibration}
\input{sections/5-evaluation}

\input{sections/6-discussion}
\input{sections/7-limitation}
\input{sections/8-conclusion}

\bibliographystyle{ACM-Reference-Format}
\bibliography{references,reference2}

\end{document}

%% file: sections/1-intro.tex

\section{Introduction}
Speech sound disorder~(SSD) is one of the most frequent communication challenges in childhood~\cite{black_communication_2015, broomfield_children_2004, shriberg_prevalence_1999, mcleod_epidemiology_2009}.
Speech sound difficulties must be understood in relation to expected development~\cite{mcleod_consonant_2018, noauthor_speech_2025-1},  and some persist beyond childhood~\cite{noauthor_speech_2025-1}.
Although the etiologies and clinical interventions of SSD vary, speech-sound practice aims to support stability and generalization of speech sound production through substantial repetition across varied linguistic contexts~\cite{noauthor_speech_2025-1, ansari_vocabulary_2025, preston_tutorial_2019, bernthal_articulation_2009, miccio_triangulating_2010, gubiani_structural_2014}.  
Conventional practice commonly relies on structured, word-level drills delivered via flashcards to elicit speech sound productions, which can disengage young children ~\cite{sugden_evaluation_2020, furlong_treating_2021, krystal_comparative_2020}. 
Gamified systems~\cite{hair_longitudinal_2021, spokeit_spokeit_2025} have also been developed to increase engagement, but word-level practice does not provide opportunities for longer, self-formulated, and spontaneous utterances, such as those used in stories and conversations. 
Story-based materials~\cite{noauthor_little_2025} can embed target sounds in connected narrative linguistic contexts~\cite{freeman_speaking_2016, wellman2011narrative, eadie_speech_2015, tambyraja_reading_2020, lefebvre_shared_2017}, but turning these materials into repeated, interactive production opportunities can still require adult preparation and prompting.
Repetitive, leveled, and engaging practice therefore remains difficult to deliver with current tools.

Outside clinical contexts, the home is one setting where adult caregivers can support children's practice~\cite{tosh_parent-implemented_2017, sugden_australian_2018, sugden_evaluation_2020, sugden_parents_2019}.
Caregivers are therefore indispensable to home practice while lacking the clinical expertise it demands~\cite{watts_pappas_parental_2016, sugden_parents_2019, dangol_i_2025}. 
In our formative study, speech-language pathologists (SLPs) cautioned that repetition without appropriate support could reinforce incorrect productions.
They also described limited visibility into what children actually produced between sessions, because caregiver reports did not preserve the production and their elicitation context.
Supporting child--caregiver practice thus requires both resources for the immediate attempt and records that can inform subsequent professional guidance.

Our formative study with 6 SLPs identified opportunities to structure repeated productions, story-based elicitation, in-session support, and production-level information for later professional review. 
A subsequent expert walkthrough with 7 SLPs used the functional prototype as a design probe to refine requirements for more precise configuration, actionable support, and selective review within clinicians' availability.
Two of these requirements pulled against each other: clinicians wanted production accuracy quantified for their own planning, and wanted no score placed in front of the child.

We propose \sysname{}, an AI-mediated role-playing story-based system in which target-bearing words recur as characters, locations, and props, eliciting productions under two response constraints while retaining recordings for later clinician review.
Cues, modeled audio, listen-back, and lexical target-registration feedback support each attempt, while quality judgments are routed to a clinician-facing dashboard rather than shown during practice.
Automated phoneme-distance scores~\cite{karhila_transparent_2019, 7319655} computed from a separate IPA-transcription model rather than the child-facing recognizer tracked recognition artifact more than child error---artifact contributed over three times the unique variance, and more than half the score's variation tracked neither---and did not preserve the ordering of severity groups; the dashboard therefore abstains rather than classifies where the score is least trustworthy.

To explore how children and caregivers used \sysname{}, we conducted a single-visit lab study with 10 child--caregiver dyads, five with a reported SSD or prior therapy.
The two modes elicited the response forms they were designed for, but the same nominal level asked caregivers for different work: reading support in one mode, formulation support in the other. 
Keeping phonemic labels off the practice screen left lexical target-registration as the only immediate signal, and at least one family read it as a verdict on pronunciation.
Families anticipated a place for the system in their routines and located the remaining demand in coordination---initiating, scheduling, and deciding what comes next---rather than correction. 
We draw implications for shared, revisable configuration, scope-bounded feedback, and support for coordination rather than correction.

This work makes the following contributions:
\begin{itemize}
    \item A clinician-involved design process---a formative study with six SLPs and an expert walkthrough with seven---that specifies requirements for configuring practice, supporting an attempt, and reviewing records, and marks which of them \sysname{} meets.

    \item \sysname{}, which binds target words to recurring story roles and varies response constraints across two modes, supporting each attempt with cues, modeled audio, listen-back, and lexical target-registration without showing quality judgments to the child; and a calibration study showing why its dashboard labels, calibrated against SLP ratings, abstain over the score range where SLP-rated tiers overlap rather than classify every production.

    \item Empirical findings from 10 child--caregiver dyads on three tensions this design surfaces: response constraint and comprehension load vary independently; a signal's evaluative weight follows its immediacy rather than its intended scope; and reducing adult correction relocates, rather than removes, the demand for adult coordination.
    
\end{itemize}



%% file: sections/2-related-work.tex
\section{Background}

\subsection{Speech-sound Practice Across Contexts and Its Hierarchical Structure}
\label{sec:background-1}
Producing speech sounds accurately is a learned skill, and non-target productions arise naturally along the way—whether a sound is acquired late in typical phonological development~\cite{mcleod_consonant_2018}, shaped by the phonology of another language a child speaks~\cite{karhila_pronunciation_2023, Amrate_Tsai_2025}, or persistently difficult due to a speech sound disorder~\cite{mcleod_consonant_2018, crowe_childrens_2020}.
What varies across these situations is the persistence and clinical significance of the errors; what remains constant is how they are addressed: repeated, structured production of target sounds with feedback, progressing through increasingly complex linguistic contexts~\cite{noauthor_speech_2025-1, ansari_vocabulary_2025, preston_tutorial_2019}.
Because this practice structure is most systematically developed in clinical SSD intervention, we review that literature as the ground for our design.

Clinical intervention for SSD targets speech production through structured practice (imitation, drills, minimal-contrast therapy) guided by SLPs~\cite{wren_systematic_2018, shi2025ai}.
In both clinical and home programs, targets are scaffolded through a developmental hierarchy~\cite{Glaspey2022Moving, rodgers2023shared}: accurate production is first established in isolation through auditory discrimination, imitation, and phonetic placement, then extended to words of increasing syllable complexity (minimal pairs, repetition, games), and finally to phrases, sentences, and conversational or narrative contexts that promote carryover to everyday communication~\cite{Storkel2022Minimal, Combiths2021Phonological, cabbage2022speech, wren_systematic_2018, Jesus2019Comparing}.
Instructional strategies that follow this progression support long-term generalization~\cite{allen_intervention_2013}.
Many digital interventions, however, still emphasize phoneme- or word-level drills with limited scaffolding into naturalistic communication~\cite{duval2017mobile, hair2018apraxia, hair_longitudinal_2021, gavcnik2018user, saul2020feasibility}, and narrative approaches that bridge levels by embedding targets in meaningful contexts~\cite{freeman_speaking_2016, noauthor_little_2025, ansari_designing_2025, wellman2011narrative} often require substantial manual preparation by SLPs or caregivers.

\subsection{Intervention Challenges: Generalization and Home Practice}
\label{sec:background-2}
While structured practice can improve in-session performance, therapy effectiveness ultimately depends on generalization across settings and communication partners~\cite{allen_intervention_2013}. 
Given the complexity of speech sound development~\cite{Combiths2021Phonological, rodgers2023shared}, generalization often requires high-frequency practice beyond the clinic to maintain consistency. 
Home-based and parent-implemented practice can increase intervention intensity and engagement and improve outcomes, particularly when supported by digital tools~\cite{Matthews2021Application, Combiths2021Phonological}. 
Critically, studies highlight that successful home practice depends on adapting activities to the child's context, providing individualized tasks, and involving caregivers in therapy planning~\cite{Leafe2024What, Leafe2025To, Leafe2025What}. 
However, many families struggle to implement activities consistently due to time constraints and limited expertise, and rising caseloads further limit SLPs' capacity for individualized support~\cite{sugden_parents_2019, blood_predicting_2002, denny_behind_2024}. 
These constraints motivate tools that sustain high-quality practice at home while maintaining continuity with clinical care~\cite{shi2025ai, Jesus2019Comparing, Matthews2021Application}.

\subsection{Digital Tools and Technologies Supporting Speech Practice}
\label{sec:background-3}
A wide array of digital tools supports speech-language therapy for children~\cite{gosnell2011apps, furlong2018mobile, du2023they, barnett2024qualitative, vaezipour2020mobile, santos2019treating, heyman2020identifying}. 
Clinician-developed applications~\cite{noauthor_little_2025, heyman2020identifying, du2023they} digitalize paper-based stimuli; Apraxia World~\cite{hair_longitudinal_2021} and SpokeIt~\cite{spokeit_spokeit_2025} apply gamification to sustain repetition. 
Ansari~\textit{et al.}~\cite{ansari_designing_2025} incorporate multisensory cues, which aligns with best practices in speech-language learning. 
Closest to \sysname{}'s architecture, Parnandi~\textit{et al.}~\cite{parnandi_development_2015} pair a child-facing practice game with automated utterance scoring and a clinician-facing web interface through which an SLP assigns exercises, reviews the individual recordings alongside their automated scores, and adapts the program. These tools particularly support phoneme and word-level practice through drill-based formats. 

Generative AI and automatic speech recognition (ASR) have further expanded the design space.
AI agents can simulate conversation partners, adapt content in real-time, and reduce the manual burden on SLPs ~\cite{du_bridging_2020, du2023they, suh_opportunities_2024, du_generative_2023, shi2025ai}.
Closest to our setting, Dangol~\textit{et al.}~\cite{dangol_i_2025} interviewed parents about home-practice challenges and developed AI-support concepts that SLPs reviewed, situating AI assistance within the informational, emotional, and practical support parents need; StoryMate~\cite{chen2025characterizing} uses LLM-supported conversation and customizable reading modes for personalized story interaction with children, parents, and educators; \sysname{} organizes the same responsiveness around repeated production of specific sounds.

ASR, however, remains unreliable precisely where practice systems need it most: assessment accuracy degrades on child speech containing non-target productions, making it difficult to distinguish production errors from recognition artifacts~\cite{7319655, tobin_automatic_2024}.
The constraint is not unique to disordered speech---CAPT research on children's L2 pronunciation reports the same degradation~\cite{Amrate_Tsai_2025}, and has explored articulatory-weighted scoring to make automated assessment more transparent~\cite{karhila_transparent_2019}.
This shared limitation across clinical and non-clinical contexts motivates treating automated scores as triage signals routed to human judgment rather than as diagnostic measures, the approach \sysname{} adopts.

How children and families act on such signals is better characterized in child--computer interaction than in the speech-technology literature.
Young children attribute intelligence and trustworthiness to voice agents~\cite{druga_hey_2017, lovato_siri_2015}, and when recognition fails, families repair by repeating, rephrasing, and raising their voices, with parents scaffolding the child's phrasing~\cite{beneteau_communication_2019}.
In sustained home use, parents mediate children's access and fold the agent into existing routines~\cite{garg_he_2020}, and co-use of digital media reshapes the parent's role rather than removing it, sometimes by crowding the parent out~\cite{hiniker_lets_2018}.
Conversational agents have also served as dialogic reading partners that question and scaffold children through a story, with and without a parent co-reading~\cite{xu_exploring_2020, xu_rosita_2023}.
StoryBuddy~\cite{zhang_storybuddy_2022} makes the parent's role an explicit design parameter in such a system: parents choose how much of the question-asking to delegate to the agent, and a parent-facing view lets them review what the child was asked and how the child answered. Its concern is the division of labor between parent and agent over a shared activity; what it does not have to handle, and \sysname{} does, is an activity in which the agent's registration of an utterance is itself read as an evaluation of the child.
These studies characterize the interaction \sysname{} places children in; none examines an activity whose purpose is repeated, monitored production of specific sounds, where the agent's registration of a word acquires an evaluative reading.

The gap \sysname{} addresses is therefore not LLM storytelling, ASR-based feedback, or a clinician dashboard, each of which exists.
It is how an implemented practice activity elicits target-bearing responses from a child, how a co-present caregiver supports those responses, and how the dyad interprets automated signals whose reliability is bounded---a combination that neither the speech-technology nor the child--agent literature has studied in a speech-sound practice context.



%% file: sections/3-design.tex
\section{Clinician-involved Design Process}
\label{sec:design}
To ground our design in a deeper understanding of speech-sound practice, we conducted a formative design study~\cite{zimmerman_research_2007} with six licensed SLPs (CCC-SLP) and an expert walkthrough~\cite{POLSON1992741} with seven SLPs.
Both protocols were approved by our institution’s IRB.

\subsection{Formative Study: Understanding Speech-Sound Practice}
The formative study asked: 
(1) how the SSD clinical workflow and challenges formed; (2) how SLPs progress practice across linguistic complexity levels; and (3) how SLPs utilize current technology for preparing therapy materials.
An interview covered existing practices and constraints; an early-prototype walkthrough elicited reactions to translating those practices into an interactive practice system.

\paragraph{Participants}
\begin{table*}[t]
\small
    \centering
    \begin{tabular}{c c c c c c c}
     \hline
         \textbf{Alias} & \textbf{Gender} & \textbf{Age} & \textbf{Work Exp.} &  \textbf{Employment Type} & \textbf{Work Setting} & \textbf{\# of SSD clients} \\  \hline
         F1 & Female & 20--29 & 1--5 years & Mainstream Primary School & Mainstream Primary School & 30+ \\  \hline
         F2 & Female & 20--29 & 1--5 years & Pediatric Outpatient Clinic & Clinics & 30+ \\  \hline
         F3 & Female & 20--29 & 1--5 years & Private practice & Clinics & 30+ \\  \hline
         F4 & Female & -- & 11--15 years & Pre-K-22 public school & Pre-K Turning 22 program & 1 -- 10 \\  \hline
         F5 & Female & 20--29 & 1--5 years & AI Services company/Private practice & Clinics/Their home & 21--30 \\  \hline
         F6 & Female & 30--39 & 1--5 years & Mainstream Primary School & Mainstream Primary School & 30+ \\  \hline
    \end{tabular}
    \caption{Formative-study participants: six licensed SLPs (F1--F6), all female and based in the United States. Ranges are self-reported; ``--''indicates not reported.}
    \label{tab:formativestudy-participants}
\end{table*}

We recruited six licensed SLPs who had provided SSD interventions, through professional referral (Table~\ref{tab:formativestudy-participants}); all were female and based in the United States.
A demographic survey adapted from Ansari~\textit{et al.}~\cite{ansari_impact_2022} characterized their work.
On 5-point scales, they reported assigning home practice relatively infrequently~(M=2.7, SD=1.51) and using technology for clinical content often~(M=4.0, SD=0.89), including chat-based LLMs to prepare materials, YouTube videos for audiovisual examples, and tablet applications for activities.
We compensated them with a USD~\$50 gift card for their participation.

\paragraph{Study Protocol}
Each hour-long session was conducted individually over Zoom.
A 30-minute contextual interview covered participants' workflow and challenges in delivering SSD interventions and assigning home practices.
Where available, participants screen-shared their current physical or digital materials, letting the researcher probe how those materials supported repetition, cueing, and progression across linguistic levels.
A researcher then walked participants through a static, clickable proof-of-concept: a two-page child-facing digital story with /r/, among the most commonly targeted sounds, embedded in the narrative.
The mock-up illustrated the proposed interaction flow and child--caregiver scenario but implemented no processing pipeline, role-play, or session review.
Participants commented on each screen, and sessions closed with questions on opportunities and risks of technology in speech-sound practice.

\paragraph{Analysis Methods}
Sessions were recorded and transcribed, then analyzed with the thematic analysis described in Section~\ref{sec:child-analysis-methods}; the research team finalized themes through discussion.
We distinguished clinicians' accounts of existing practice, which ground the findings below, from reactions elicited by the prototype walkthrough, which we use only to trace how those findings were translated into specific design decisions.

\subsection{Formative Findings}
\label{sec:formative-finding}
Clinicians converged on a view of speech-sound practice as dose-dependent: \formative{6} highlighted the repetition, \textit{``Speech is different from language. Speech is all about repetition''}.
\formative{4} characterized the evidence-based target as \textit{``high frequency, lower dose [\ldots] the ultimate is 5 to 10 minutes a day, similar to any type of exercise''} yet scheduled delivery is structurally unable to supply: \textit{``speech is offered at half hour intervals for scheduling purposes [\ldots] So it's not the optimal way to really treat those sounds''}.
\formative{1} highlighted \textit{``30 minutes a week isn't going to be enough''}.
Between-session practice was therefore necessary, and our analysis identified four constraints that become consequential when practice extends beyond the clinic.

\subsubsection{Practice intensity depends on structuring repeated opportunities for production}
\label{formative:story}
Clinicians treated repeated production as central to practice: \formative{6} summarized the goal as, \textit{``we want to get their response as many as we can''}.
Because repetition alone did not hold attention,
clinicians interleaved productions with board games and matching activities, pausing each turn for \textit{``sets of, like, 3 to 5 trials''} before \textit{``the kid gets rewarded with a turn''}, \formative{3} mentioned.
The game sustains engagement, but the clinician still supplies every practice occasion.
\formative{2} described the resulting cost, \textit{``it reduces the number of times I can [\ldots] get the word in practice''}.
\formative{2} also suggested the trade-off is not inevitable: playing Chutes and Ladders with a child working on /s/, landing on a `slide' became a production occasion, \textit{``embedded inside the game, not something that's separate''}, but achieving this requires deliberate matching.
Engagement and practice density therefore diverged when play and production separated, 
which matters most outside the clinic, where no SLP structures each turn. 
Narrative materials place targets in connected contexts, but clinicians complained existing versions were too passive: \formative{6} called them \textit{``slow''} and \textit{``not intensive''}.
Home practice therefore required target-bearing opportunities built into the interaction itself, so that elicitation did not depend on an adult prompting each production.

\subsubsection{Practice becomes harder to structure as speech becomes more connected}
\label{formative:levels}
Clinicians described practice as a progression: \formative{4} described, \textit{``it's really kind of like a stepladder [\ldots] sound, syllable, word [\ldots] phrase, sentence, conversation''} in which accuracy at one step does not carry to the next.
One reason is co-articulation, as in \formative{4}'s example: a stable /d/ in isolation, then \textit{``dog,''} then \textit{``brown dog, the N then impacts the D from brown into dog''}, concluding that \textit{``the more complex the sentence, the conversation, the more influence other sounds have over that sound.''}
Conversation was consistently the hardest level because articulation competes with formulation: \formative{3} identified the difficulty in competing cognitive demands in articulation and sentence formulation: it is \textit{``more casual, and it's easier to kind of forget about articulating that sound.''}
\formative{1} employs role-playing, which is \textit{``more unstructured''}, for children who \textit{``have already made a lot of progress''}, since in free talk a child \textit{``can easily mispronounce''} unnoticed.
In school-based practice, this transition can coincide with reduced direct support: \formative{6} remarked that \textit{``conversation level is not the speech session that we normally do [\ldots] you already got this skill, I want you to carry this skill into your daily life and use that in the conversation.''}
The distinction is also visible outside the clinic. 
Clinicians, such as \formative{4}, see progress at the word-level while parents report that, \textit{``in conversation, he still says GOG for dog''}; therefore, high-level practice is difficult regardless of lower-level performance.
With younger children, an open conversational prompt may be hard to elicit intended form, ``\textit{they'll give you one word, and that's it,''} which \textit{``defeats the purpose of doing conversational level.''}
Higher-level practice requires scaffolding that prompts children to produce increasingly connected speech without collapsing the activity back to isolated-word drills.

\subsubsection{Caregiver preparation did not ensure support at the moment of production}
\label{formative:caregiver}
Consistent with prior work on caregivers' uncertainty about modeling and cueing~\cite{sugden_parents_2019, dangol_i_2025, watts_pappas_parental_2016}, SLPs described substantial efforts to prepare caregivers.
Worksheets, word lists, invitations to observe sessions, and exercises attached to daily routines.
\formative{2}, for example, suggested five to ten repetitions during the drive to school or brushing teeth.

These strategies, however, did not ensure that support was available when the child produced a target.
\formative{3} noted that many caregivers' implementation was \textit{``not always consistent''} because they have their own work; \formative{2} remarked routine-based plans \textit{``don't always work''}, and \formative{5} treated it as \textit{``unrealistic to think that they're going to hear what I'm saying to them after the session.''}
The limitation was not simply a lack of caregiver knowledge or willingness. 
\formative{4} addressed the real obstacles, \textit{``it's not that parents are not equipped, because they are''}, but \textit{``that's not what parents are for. Parents are there to take care of their children and not really instruct them.''}
\formative{4} shared their colleagues' case whose children nonetheless \textit{``don't want to work with them, even though they're extremely qualified''}.
Transferring instructions to caregivers therefore did not reproduce the clinician--child interaction.
Caregivers still had to remember the cue, decide when to intervene, and ask the child to repeat.
This gap was clinically consequential, as \formative{4} warned, ``\textit{A lot of times the homework if you're practicing the incorrect way and you don't have somebody saying nope, [\ldots] they continue to perpetuate the wrong movement of the articulator. So giving homework if they haven't solidified that sound is sometimes like five steps forward, three steps back}''.
These accounts motivate point-of-practice support that supplements, rather than replaces, caregiver involvement.

\subsubsection{SLPs need production-level visibility into home practice}
\label{formative:data}
A separate problem arises after the attempt: the SLP responsible for the next clinical decision may have little direct evidence of what occurred.
Caregiver reports can indicate whether practice happened or whether speech sounded different at home, but discontinuity is described by \formative{5}, ``\textit{They might be doing really well with me in the treatment room, but it doesn't sound the same when they're outside and practicing at home}''.
The missing information was not simply whether a child completed an activity. 
Clinicians described progress in terms of target, linguistic level, amount of cueing, and consistency of production.
\formative{6} shared that they decided to move from phrase to sentence level if the sound is produced with limited cueing, and treated such judgments as a routine: \textit{``You don't do a standardized assessment all the time, but you do an informative assessment all the time.''}
%
A completion count or caregiver summary alone therefore cannot show what the child was asked to produce, what they actually produced, or in what context, leaving SLPs with a limited basis for relating between-session practice to the next target or level, or for inspecting whether a difficulty emerges only in connected speech.
These accounts establish a requirement for production-level visibility, not for a particular automated judgment.

\subsection{Expert Walkthrough: Examining Design Assumptions}
\label{sec:expert}
We translated these four findings into a functional prototype: target-bearing words became recurring characters, places, and props within role-play; Word/Sentence and Conversational Modes varied response constraints; modeled audio, mouth-picture cues, listen-back, and retry supported the moment of production; and a dashboard retained recordings with their elicitation context under provisional labels. 
An expert walkthrough with seven licensed SLPs then examined whether these choices provided appropriate practice demands, actionable support during an attempt, and records clinicians could review within their workloads.
We treated it as generative rather than summative: SLPs' reactions were a source of requirements, not a verdict on the system.
It asked (1) what form of expert oversight is practical given SLPs' existing workflows and caseloads, and (2) where \sysname{}'s practice components meet clinical requirements and where they fall short.

\paragraph{Participants}
\begin{table*}[t]
\small
    \centering
    \begin{tabular}{c c c c c c c c }
     \hline
         \textbf{Alias} & \textbf{Gender} & \textbf{Age} & \textbf{Degree} & \textbf{Work Exp.} &  \textbf{Employment Type} & \textbf{Work Setting} & \textbf{\# of SSD clients} \\  \hline
         E1 & Female & 50--59 & Ph.D & 21+ years & Self-employment & Client's home & 30+ \\  \hline
         E2 & Female & 50--59 & Ph.D & 21+ years & Pre-K/Kindergarten & Mainstream Primary School & 30+ \\  \hline
         E3 & Female & 60+ & Ph.D & 21+ years & Self-employment & Their Home/Office & 30+ \\  \hline
         E4 & Female & 40--49 & Ph.D & 11--15 years & University/College & University/College & 1--10 \\  \hline
         E5 & Female & 20--29 & M.S. & 1--5 years & Hospital & Hospital & 21--30 \\  \hline
         E6 & Female & 50--59 & Ph.D & 21+ years & University/College & Literacy Based Preschool & 30+ \\  \hline
         E7 & Female & 20--29 & Ph.D & 1--5 years & Self-employment & Private Practice & 1--10 \\  \hline
    \end{tabular}
    \caption{Expert-walkthrough participants: seven licensed SLPs (E1--E7), all female and based in the United States; E5 and E7 also joined the formative study.}
    \label{tab:review-participant}
\end{table*}

Seven licensed SLPs~(Table~\ref{tab:review-participant}) took part; two (\expert{5}, \expert{7}) had also participated in the formative study, and the remaining five were recruited by cold email.
All SLPs are female and based in the United States.
On a 7-point Likert scale (1 = Never, 7 = Very Often), they reported neutral development/hand-over of home practice (M=4.0, SD=1.91).
They rated their technology use for clinical content as neutral~(M=4.0, SD=0.82).
Participants were compensated with a USD~\$60 gift card for their participation.

\paragraph{Protocol}
We prepared two story configurations (/l, r/ and /s, r/) with target words suggested by \sysname{}.
After consent and a demographic questionnaire~\cite{ansari_impact_2022}, a researcher operated the prototype over screen-sharing, guiding each SLP through a usage scenario covering both interaction modes and the clinician-facing dashboard while asking them to take actions, revisit features, and comment throughout.
Participants then completed a 7-point Technology Acceptance Model (TAM) survey~\cite{venkatesh_technology_2008}, and a survey on clinical relevance, practice intensity, balance, and educational impact.

\paragraph{Analysis}
Sessions were recorded, transcribed, and analyzed with the thematic-analysis procedure described in Section~\ref{sec:child-analysis-methods}. 
The survey responses are reported descriptively.

\subsection{Expert Walkthrough Findings}
\label{sec:expert-findings}
Post-walkthrough ratings were favorable: participants anticipated integrating the activities into practice~(M=6.0, SD=1.15) and perceived alignment with current intervention~(M=6.1, SD=1.46).
These ratings reflected impressions formed during a researcher-led demonstration, rather than independent use or therapeutic outcomes.
The interviews qualified this reception with requirements the demonstrated features did not fully address: specifying an appropriate practice demand, supporting families during an attempt, and making recorded productions reviewable without treating automated scores as clinical judgments.

\subsubsection{Clinical configurability must extend beyond selecting a target sound}
\label{expert:configuration}
SLPs saw value eliciting speech within a story as motivating practice contexts~(M=6.6, SD=0.53). 
\expert{5}, for example, valued children creating their own sentences rather than copying a supplied sentence.
Their comments on the example stories showed, however, that selecting target sounds and a mode did not specify what a child would be asked to practice.

\paragraph{Word selection required more than phoneme coverage.}
When the prototype offered \textit{squirrel}, \expert{2} noted that its combination of /r/, /l/, and an initial cluster made it especially difficult for many of her clients.
\expert{1} similarly questioned \textit{lizard} for a child struggling with both /l/ and /r/, and requested more specific constraints, such as \textit{``L, word initial, single syllables''} or \textit{``Medial position, multisyllabic words only''}.
Familiarity introduced another consideration: \expert{4} contrasting the potentially unfamiliar \textit{sample} with \textit{strawberry}.
A word could thus contain the requested sound while introducing other production demands or unfamiliar vocabulary---differences within a nominal practice level.

\paragraph{Precise control should not require manual assembly.}
\expert{1} wanted to \textit{``control precisely''} what entered the practice material, but doubted if she would have time to do so manually, valuing the convenience of \textit{``plug and play''}: selecting an approach and relevant sound constraints, then having the system generate an appropriate word set.
The request was therefore not for more editable fields, but for automation that could carry out a specified plan while leaving its contents inspectable and changeable.

\paragraph{Modeled exposure and child production required separate control.}
The story also bundled together hearing target sounds and producing them.
\expert{1} questioned why so many sounds were spoken by the application rather than the child, and asked whether this auditory exposure could be reduced for children who did not need it.
\expert{3} instead wanted children to listen to individual target words before progressing to sentence-level material, rather than optional exposure.
\expert{5} requested predictable sets of five or ten attempts to make the practice count clear for reporting.
These requests did not converge on more repetition.
They called for distinguishing how much modeled input a child heard from how many productions the activity requested, extending configuration beyond a sound-and-mode choice to the composition, familiarity, and amount of practice.

\subsubsection{Automated signals could prompt action but not settle production quality}
\label{expert:in-session-support}
\label{expert:review}
Clinicians moderately agreed that the represented cues were adequate~(M=5.5, SD=1.52; $N{=}6$, with \expert{1}'s response to this item excluded) and rated the dashboard very useful for tracking~(M=6.43, SD=0.79) and informing the children's practice with audio clips and transcripts~(M=6.71, SD=0.49).
They nonetheless distinguished a request for another attempt from guidance on what to change, access to records from the ability to incorporate them routinely, and the convenience of a compact summary from the credibility of the judgment it appeared to convey.


\paragraph{A retry request did not necessarily identify what to change.}
During \expert{4}'s walkthrough, the system requested further attempts at \textit{seaside} although she judged the researcher's production to be correct.
She anticipated that such a discrepancy could confuse a child who had followed the articulatory instructions, and attributed it to possible accent or stress sensitivity, though the walkthrough did not establish its technical cause: ``\textit{The treatment aims not to focus on the accent, stress, or intonation. Our focus is just to produce sound correctly}''.
Separately, \expert{6} questioned feedback that flagged a problem without explaining \textit{``why it's wrong''}, noting that productions could become closer to the target without becoming fully accurate.
These accounts distinguished eliciting another attempt from helping someone improve that attempt.
They left open how a child or caregiver should interpret an unregistered target, decide whether a change in production was needed, and determine what support to use next.

\paragraph{Capturing records did not make review routine.}
\expert{5} valued being able to return to the child's actual recording, describing the dashboard's listen-again capability as \textit{``extremely helpful''}.
\expert{1} likewise appreciated recordings and transcriptions, but immediately qualified their practical value: \textit{``I don't have time to go back and check their homework''}.
She questioned whether she could review results across a large caseload and proposed a weekly summary email rather than having to inspect each child's records separately.
Thus, moving review outside the practice session did not remove its workload; it created a need to make re-engagement brief and selective.


\paragraph{Compact labels could not settle production quality.}
\expert{7} cautioned against relying completely on automated scoring, particularly for /r/, where she described difficulty distinguishing an accurate production from a near-target one.
A recognition mismatch led \expert{5} to ask whether clinicians could manually count an unregistered production of \textit{library}, while \expert{3} raised confidentiality and access requirements for retained voice recordings.
\expert{7} also valued keeping scores out of the ongoing child-facing activity, anticipating that visible scoring could undermine confidence in further attempts.
Together, these accounts supported review through inspectable records and provisional summaries, while leaving both the feasibility of routine review and the reliability of automated interpretation open.

\subsection{Design Responses and Remaining Questions}
\label{sec:design-response}

\begin{table*}[t]\small
\caption{Requirements from the clinician-involved design process, \sysname{}'s response, and what this paper establishes about each.}
\label{tab:requirements}
\begin{tabularx}{\textwidth}{@{}
    >{\raggedright\arraybackslash}p{0.23\textwidth}
    >{\raggedright\arraybackslash}p{0.28\textwidth}
    >{\raggedright\arraybackslash}X
@{}}
\toprule
\textbf{Requirement} & \textbf{Response} & \textbf{Status and evidence} \\
\midrule
Repeated production built into the activity & Targets bound to recurring characters, places, props & Retained. Design-time capacity; realized opportunities. Not compared to drill \\
Response less constrained toward connected speech & Word/Sentence and Conversational Modes & Retained. Response forms; Modes self-selected \\
Support at the point of production & Cue panel, placement description, gesture, modeled audio, listen-back & Retained. Support acts; interpretation \\
Production-level visibility for later review & Dashboard retaining audio, target, prompt, transcripts, label & Retained. Expert ratings. No clinician reviewed records from actual practice \\
Configuration beyond sound and mode & Sound, syllable complexity, vocabulary configurable & \textbf{Partly open}. Position and cluster not implemented; vocabulary mismatch \\
Modeled exposure and production count controlled separately & --- & \textbf{Open} \\
A retry request should indicate what to change & Registration retry, capped at two, continuation permitted & \textbf{Open}. Read as a pronunciation verdict \\
Selective review without authoritative scoring & Three SLP-calibrated labels with an abstention band & Revised. Calibrated on STAR; not validated on study recordings \\
Confidentiality of children's voice data & --- & \textbf{Open}. Study-level agreements only \\
\bottomrule
\end{tabularx}\end{table*}

Table~\ref{tab:requirements} summarizes what the two clinician studies required, what \sysname{} does in response, and what the present paper does and does not establish.
Two of the retained responses rest on clinicians' anticipations rather than observed use.
One requirement was revised after the walkthrough: we replaced the provisional categorization with three labels calibrated against independent SLP ratings (Section~\ref{sec:calibrated-labels}), whose middle category marks a range where automated classification is untrustworthy rather than a detected borderline production.
That revision concerned the clinician-facing pipeline only and therefore did not address the retry-interpretation problem the walkthrough exposed.

Three requirements remain open, and two of them---separating modeled exposure from production count, and telling a child what to change rather than only that something did not register---describe capabilities we considered and did not build.
Modeled-exposure control would have required a second configuration axis whose interaction with linguistic level we could not specify from the walkthrough.
Actionable retry feedback would have required the system to make a claim about production quality at the moment of practice, which Section~\ref{sec:scoring} shows the score cannot support.
The third, confidentiality of children's voice data, is handled by study-level agreements only.
We name them here because they bound what the child--caregiver study could observe.
The system follows in Section~\ref{sec:sys} and the calibration of its dashboard labels in Section~\ref{sec:scoring}.

%% file: sections/4-system.tex
\section{\sysname{}}
\label{sec:sys}

\subsection{Interaction Overview}
\label{sec:system-overview}
\begin{figure*}
    \centering
    \includegraphics[width=1\linewidth]{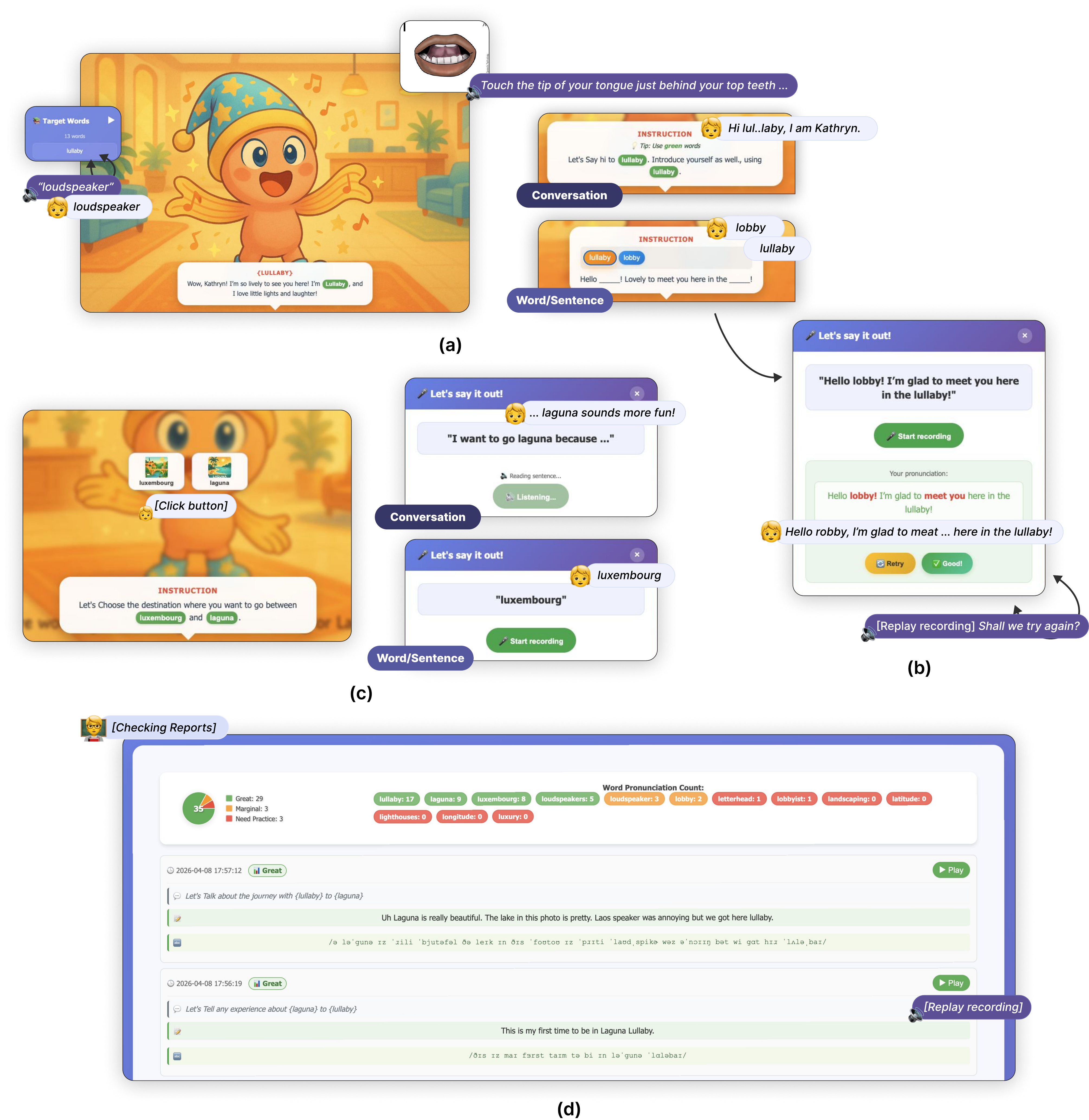}
    \caption{\textbf{Interaction overview of \sysname{}.}
    (a) During story play, target words, modeled audio, and articulatory cues remain available while Word/Sentence and Conversational Modes elicit responses with different levels of constraint.
    (b) After a recording, children can listen back to their production and receive ASR-based target-registration feedback, with an option to retry or continue.
    (c) Story interactions, such as choosing a destination, create target-bearing production opportunities: Word/Sentence Mode supplies a more constrained response, whereas Conversational Mode asks the child to formulate a response using the target vocabulary.
    (d) The clinician-facing dashboard summarizes practice and preserves individual recordings with their elicitation prompts, recognized and phoneme-level transcripts, and dashboard labels for later review.}
    \label{fig:main}
\end{figure*}

\sysname{} combines child--caregiver role-playing practice with a separate clinician-facing view of recorded productions, supporting preparation of target-bearing materials, interaction during practice, and later inspection of the resulting records.
Figure~\ref{fig:main} summarizes \sysname{}'s interaction flow across four parts. 

\subsubsection{Configuring and Eliciting Target-Bearing Practice}
\label{design:repetition}
Section~\ref{formative:story} and~\ref{formative:levels} identify a coupled requirement: repeated production opportunities must be built into the activity, while the child's response should become less constrained as practice moves toward connected speech.
\sysname{} combines the explicit lexical constraints of Mad-Lib-style prompts with the freer response structure of role-play.

\paragraph{Clinician configuration}
Under the intended workflow, an SLP selects a pre-authored story template and configures the child's target sound or blend, desired syllable complexity, and target vocabulary.
\sysname{} can suggest candidate words, which the SLP can retain or replace with child-specific words. 
The selected words are bound to recurring story roles, including characters, locations, destinations, and props.

\paragraph{Target-bearing opportunities}
Each template defines lexical slots and interaction nodes before dialogue is generated.
Designated choice and fill-in nodes request a target-bearing response, and the child's choice can determine the next story turn~(Figure~\ref{fig:main}(c)).
A sidebar alongside the story~(Figure~\ref{fig:main}(a)) provides optional opportunities to hear and practice target words without changing narrative state.
This structure fixes a minimum number of target-bearing opportunities without depending on the automated system to introduce the vocabulary spontaneously, and binding targets to recurring story roles also places them in different neighboring-sound contexts---a variation \formative{4} identified as consequential, since \textit{``red rainbow''} is harder than \textit{``yes rainbow''} for the same target sound.
This arrangement specifies where production is requested, not a guaranteed number of completed or accurately pronounced productions.

\paragraph{Varying response constraints within a shared story}
Both modes use the same scene structure and configured target words, but differ in how the child's turn is elicited, as shown (Figure~\ref{fig:main}(a,c)).
In \textit{Word/Sentence Mode}, the child supplies a target word for a story blank and is then prompted to produce the completed sentence.
In \textit{Conversational Mode}, the child instead formulates a response to a character using one or more suggested target words without being given the full sentence to reproduce.
Figure~\ref{fig:main}(c) illustrates this contrast for the same story interaction.
These are elicitation formats rather than measures of the child's realized linguistic level.
The target-word check does not establish that the response is a complete sentence or connected speech, and recordings are retained regardless of the response form.

\subsubsection{Supporting the Production with Child-Facing Cues and Feedback}
\label{design:cueing-feedback}
Section~\ref{formative:caregiver} motivates support at the point of practice.
\sysname{} does not remove caregiver involvement; instead, it provides cues and immediate feedback directly within the child-facing interaction, reducing reliance on caregivers as the sole source of modeling and evaluation. 

\paragraph{Cues and modeled exposure}
For each configured target sound, a cue panel beside the story (Figure~\ref{fig:main}(a)) provides a mouth-shape picture with an isolated sound model, a brief placement description, and a gesture cue.
Presenting them in the interface reduces what caregivers must recall and demonstrate from memory~\cite{rodriguez_prelingual_2012, darejeh_speech_2019, rusiewicz_effect_2017}.
The target-word sidebar lets children hear modeled word productions and record additional attempts, and configured vocabulary recurs in character dialogue, providing repeated modeled exposure without pausing the narrative.

\paragraph{Immediate feedback after production}
After a recording, the interface offers \textit{listen-back} and checks whether the targets specified for the interaction appeared in the recognized text (Figure~\ref{fig:main}(b)).
We call this lexical match \textit{target-registration}, not an assessment of articulatory accuracy.
\textit{Word/Sentence Mode} displays colored target-word markings and offers a retry when required targets are not registered.
\textit{Conversational Mode} uses the same lexical check to request another response containing the unregistered targets.
This design follows the walkthrough, where clinicians cautioned that a visible quality judgment could undermine a child's confidence in further attempts (Section~\ref{expert:review}); and the score itself cannot support them~(Section~\ref{sec:scoring}).
An unregistered target may reflect an omitted word, a production difference, or recognition failure; the match state does not distinguish these causes.
After two unsuccessful matches, the interface permits the child to continue the story and the recording and its context are retained whether the target is registered, another attempt is made, or the interaction advances without a match.

\paragraph{Separate decisions, separate feedback pipelines}
The two interfaces therefore expose different automated decisions:
child-facing target registration guides the immediate interaction (Figure~\ref{fig:main}(b)), whereas the dashboard (Figure~\ref{fig:main}(d)) presents labels derived from a separate phoneme-distance score for later professional interpretation (Section~\ref{sec:scoring}).
Because SLPs wanted production-level information for later review while cautioning against exposing evaluative scores during children's ongoing practice~(Section~\ref{expert:review}), dashboard labels are computed for the clinician-facing view only.
This placement is also consistent with motor-learning accounts of speech practice, in which summary feedback delivered after a set of trials supports retention while immediate high-frequency feedback aids momentary performance at its expense~\cite{maas2008principles}.
The separation concerns what each interface computes and displays; it does not establish how families interpret the markings and prompts that remain.

\subsubsection{Reviewing Recorded Productions and Their Context}
\label{design:dashboard}
Finding~\ref{formative:data} calls for production-level visibility, but not continuous professional supervision. 
\sysname{} therefore preserves each attempt and presents the resulting records through the clinician-facing dashboard~(Figure~\ref{fig:main}(d)) at two granularities: an overview of practice frequency by target word, session-level trends, and the distribution of dashboard labels to identify patterns before opening individual records.
The trial-level view retains the recording, expected target, interaction prompt, recognized and phoneme-level transcript, and dashboard label for selective review.

\subsection{Technical Pipeline and Implementation}
\label{sec:tech-pipeline}
\begin{figure*}
    \centering
    \includegraphics[width=0.8\linewidth]{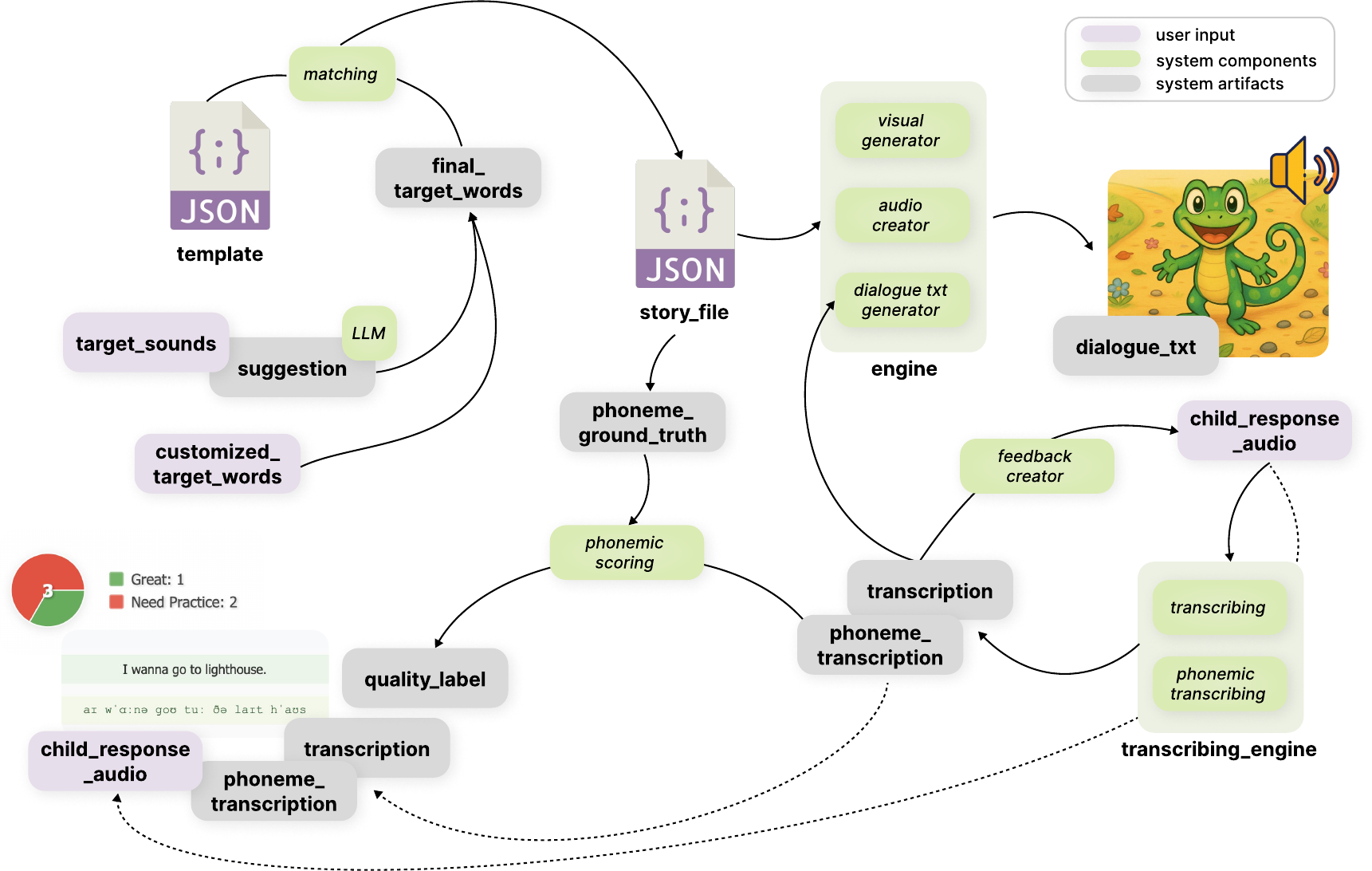}
    \caption{Technical pipeline of \sysname{}. 
    Target sounds and customized target words are combined with an authored JSON template to instantiate the story and its required target words.
    The story engine then generates dialogue, audio, and visual assets for the child-facing interaction.
    During practice, recorded child speech follows two separate processing paths: lexical transcription is used to determine whether required target words were registered and to construct child-facing feedback, while phonemic transcription and scoring produce quality labels and review artifacts for the clinician-facing dashboard.
    Colors distinguish user inputs, system components, and system artifacts.}
    \label{fig:tech-pipeline}
\end{figure*}

Figure~\ref{fig:tech-pipeline} depicts a data flow from configuration to dashboard.

\subsubsection{Template-Constrained Story Generation}
\label{sec:story-pipeline}
A three-layer pipeline maintains a predictable density of target phonemes while preserving generative flexibility.

\paragraph{Layer 1: authored narrative template}
Each story is defined by a JSON template that specifies a fixed scene graph: introduction (character meets the child), conflict (an antagonist appears), and resolution (the child and the character overcome the challenge and reach the destination).
The template encodes scene-level metadata, including characters, locations, available interactions, the turn-taking sequence between characters and the child, and branching points; branches such as destination selection are conditional edges in the graph, ensuring that every narrative path is reviewed for age-appropriateness before deployment.
 
\paragraph{Layer 2: configuration and prompt construction}
Target words containing the child's target phonemes are filled into parameterized slots such as \texttt{\{character1\}}, \texttt{\{place1\}}, \texttt{\{destination1\}}, and \texttt{\{prop\}}; for example, an /r/ configuration might use ``Rabbit,'' ``River Road,'' and ``rainbow.''
Because these slot values recur across the story, this layer guarantees a minimum target-phoneme density without relying on the language model to introduce target vocabulary spontaneously.
Across our templates, this design embeds 7--13 target words per session, comprising 6--7 child turns and additional interactions.

\paragraph{Layer 3: runtime dialogue generation}
At runtime, the language model generates character dialogue turn by turn, conditioned on the scene metadata, the instantiated slot values, and a system prompt enforcing vocabulary level for the specified age range, sentence length, and inclusion of target words.
Because the narrative frame and slot values are fixed, the model's freedom is bounded to \textit{how} dialogue is expressed---tone, phrasing, and conversational flow---rather than \textit{what} the story is about.


\subsubsection{Child-Facing Feedback Pipeline}
Child responses are transcribed by OpenAI gpt-4o-mini-transcribe, and matched, case-insensitively without stemming (the targets are mainly nouns), against the target words required by the current interaction node; a target counts as registered when its surface form appears in the recognized text.
Registration is therefore a lexical decision computed from a single transcription hypothesis and does not consult the phonemic scoring pipeline described (Section~\ref{sec:scoring}), following the requirement identified in Section~\ref{expert:review}.
Word/Sentence Mode renders this state as colored target-word markings; Conversational Mode uses the missing targets and current scene context to construct an encouraging retry prompt, voiced through text-to-speech, and the child can replay the original recording before attempting the turn again.
The audio, recognized transcript, required targets, interaction mode, and prompt context are stored for later review.

\subsubsection{Implementation Details}
\sysname{} is implemented as a web application with React/TypeScript and a Python FastAPI backend with Socket.IO supporting interactive story turns. 
OpenAI's GPT-4.1-mini generates character dialogue at runtime, GPT-4o-mini-TTS produces character and feedback audio, and GPT-Image 2.0 produces the pre-session character and scene assets.
Firebase Storage stores audio recordings and generated image assets.
Firestore stores structured story configurations, session metadata, interaction prompts, and transcripts. 

%% file: sections/5-calibration.tex
\section{Calibrating the Dashboard Labels}
\label{sec:scoring}
The walkthrough prototype grouped raw phoneme-distance scores into four provisional tiers using researcher-selected boundaries.
Clinicians valued compact summaries but questioned their authority as judgments of production quality (Section~\ref{expert:review}).
We therefore examined the score's error sources and calibrated the dashboard labels against independent SLP ratings.
This analysis characterizes the basis and limits of the labels; it does not evaluate child-facing target-registration or establish savings in professional review time.

\subsection{Scoring Method}
\sysname{} computes a \textit{phone feature error rate (PFER)} using PanPhon's feature-based Hamming edit distance~\cite{mortensen_panphon_2016}, which assigns similarity-based costs to phoneme substitutions.
Distances are divided by the length of the longer segment sequence; the prototype applies a common threshold pair to word- and sentence-level recordings, and length normalization alone does not establish equivalent calibration across these production types.
The target phoneme sequence is derived from the expected English text with the CMU Pronouncing Dictionary~\cite{noauthor_cmu_nodate}.
Before any distance is computed, all three strings---target, expert transcription, and model output---pass through one normalization step that strips stress and length marks and folds notational variants onto the forms PanPhon parses: ASCII look-alikes (\texttt{g}~$\rightarrow$~\textipa{g}, \textipa{\textteshlig}~$\rightarrow$~\textipa{tS}), r-colored vowels (\textipa{\textrhookschwa}~$\rightarrow$~\textipa{@r}, \textipa{\textrhookrevepsilon}~$\rightarrow$~\textipa{3r}), and the broad/narrow rhotic choice (\textipa{\*r}~$\rightarrow$~\textipa{r})~\cite{ipa1999handbook, ladefoged2015course}.
Audio is transcribed to IPA by Gemini~3.7~Flash\footnote{https://blog.google/innovation-and-ai/models-and-research/gemini-models/introducing-gemini-3-7-flash/}, which produces the model phoneme sequence for every distance reported below.
Five target strings encode pronunciation alternatives that the scoring pipeline concatenates into a single sequence rather than selecting between, which inflates their distances; we retain that inherited behavior so the measure is uniform across productions, and flag it as a limitation.

\subsection{Why Raw Scores Are Insufficient as Indicators}
\label{sec:error-decomposition}

\begin{table*}[t]
    \small
    \centering
    \begin{tabular}{@{}lrccccc@{}}
    \toprule
    \textbf{Reference-error group} & \textbf{$N$} & \textbf{Child Error} & \textbf{ASR Artifact} & \textbf{System Score} & \textbf{System / Child} & \textbf{Art.~$>$~Child} \\
    \midrule
    G1 ($\leq$0.059)   & 41 & 0.031 & 0.112 & 0.110 & 3.54$\times$ & 61.0\% \\ \hline
    G2 (0.059--0.160)  & 40 & 0.100 & 0.242 & 0.249 & 2.49$\times$ & 65.0\% \\ \hline
    G3 (0.160--0.268)  & 40 & 0.220 & 0.265 & 0.204 & 0.93$\times$ & 65.0\% \\ \hline
    G4 ($>$0.268)      & 41 & 0.387 & 0.243 & 0.326 & 0.84$\times$ & 19.5\% \\
    \bottomrule
    \end{tabular}
    \caption{Error decomposition by expert-transcribed error group (disordered word-level, $N{=}162$ productions from seven speakers). The system score overstates child error by $3.54\times$ in the lowest group and understates it in the two highest, and its group means are not monotone in child error: G2 scores higher than G3 despite lower actual error. Group boundaries are empirical quartiles of the recomputed child-error scores, with upper bounds included; ratios divide group means. Three productions have zero child error and enter no per-production ratio.}
    \label{tab:phoneme}
\end{table*}

Off-the-shelf ASR models, trained predominantly on adult speech, introduce systematic errors on children's speech, compounding when the child has a speech sound disorder.
To characterize this effect in our own setting, we decomposed the score on the 162 disordered word-level productions for which expert phonemic transcriptions are available in the STAR Database~\cite{noauthor_star_2025} into \textit{child error} (target vs.\ expert transcription; $M{=}0.185$, $SD{=}0.149$) and \textit{ASR artifact} (expert transcription vs.\ model prediction; $M{=}0.215$, $SD{=}0.165$).
The system score ($M{=}0.222$, $SD{=}0.184$) exceeds child error in $53.7\%$ of productions, with a ratio of means of $1.20{\times}$.

The mismatch is systematic rather than uniform.
Grouping productions by their expert-transcribed error, the system score overstates child error by $3.54{\times}$ in the lowest group and understates it at $0.84{\times}$ in the highest (Table~\ref{tab:phoneme}).
Two consequences follow. The spread between the extreme groups shrinks from $0.356$ in child error to $0.216$ in the system score, so the automated measure compresses precisely the signal a reviewer would want it to preserve.
More seriously, it does not even preserve the ordering: productions in the second group (child error $0.100$) receive a higher mean system score ($0.249$) than those in the third (child error $0.220$, system score $0.204$).
A raw distance therefore cannot be read as a severity ranking, let alone as a verdict on a single production.

In a regression of system score on the two components, ASR artifact alone explains $37.7\%$ of system-score variance and child error alone $19.0\%$; together they reach $46.2\%$.
Partitioning the joint model, artifact contributes $27.2\%$ uniquely against $8.5\%$ for child error, with $10.5\%$ shared and $53.8\%$ unexplained by either.\footnote{Unique contributions are the joint $R^2$ minus the other predictor's single-predictor $R^2$; shared is the sum of single-predictor $R^2$ values minus the joint $R^2$. These are descriptive associations, not a causal decomposition: the three pairwise distances do not sum. An alternative allocation of shared variance is reported in the supplementary material.}
Artifact thus accounts for roughly three times the unique variance that child error does, and more than half the score's variation tracks neither component.
Automated scoring is least reliable where clinical judgment is hardest, on productions close to target.


\begin{figure*}[t]
    \centering
    \includegraphics[width=\linewidth]{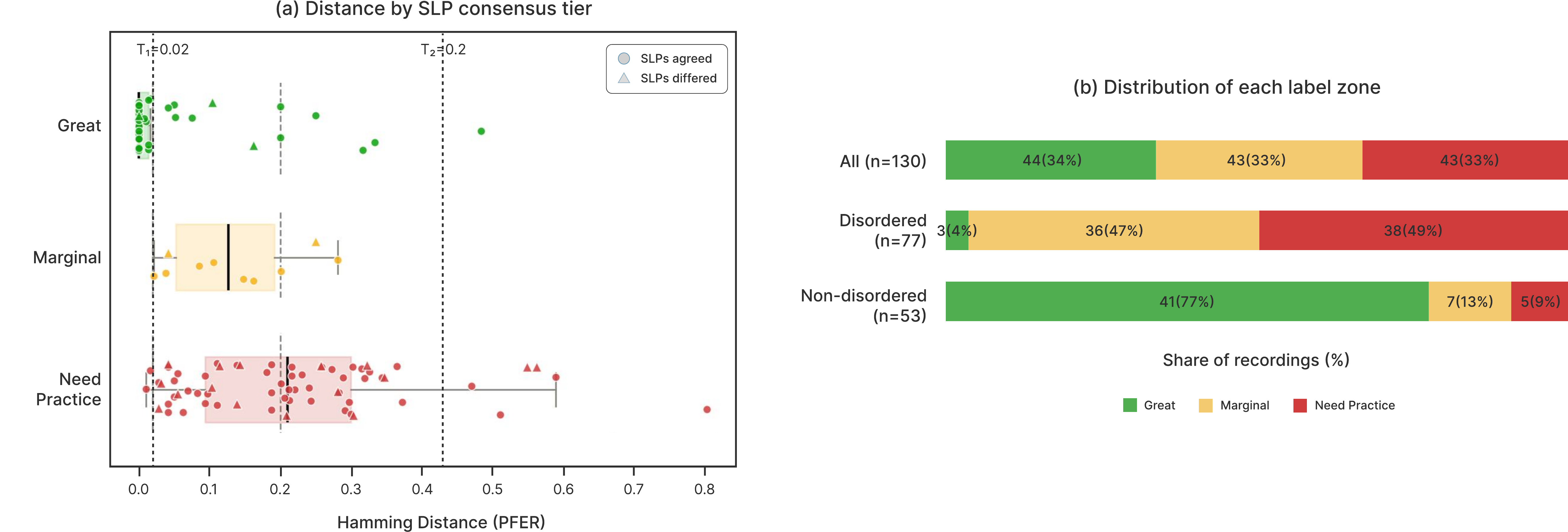}
    \caption{SLP calibration analysis ($N=130$). (a) Each point represents a recording's normalized phoneme-feature distance, grouped by SLP consensus tier. Marker shape distinguishes recordings on which the two SLPs agreed or differed. Dashed lines indicate the dual-threshold boundaries at 0.02 and 0.20. (b) Distribution of automated dashboard labels across the full calibration set and its disordered and non-disordered subsets.
    }
    \label{fig:calibration}
\end{figure*}

\subsection{SLP-Calibrated Dashboard Labels}
\label{sec:calibrated-labels}
Two licensed SLPs independently rated a 130-recording calibration set from STAR: 47 disordered word-level, 53 non-disordered word-level, and 30 disordered sentence-level recordings.
Under the three-tier evaluation---\textit{Great}, \textit{Marginal}, \textit{Need Practice}---agreement was substantial (Cohen's weighted $\kappa_w{=}0.820$, exact agreement${=}83.8\%$); remaining disagreements concentrated in the middle tier.

Because recognition artifacts can dominate relatively small production differences near the target, well-produced recordings could receive high distances and poorly produced recordings low ones, producing substantial overlap across SLP-rated tiers:
distances for \textit{Great} productions span $0.00$--$0.48$ and those for \textit{Need Practice} span $0.01$--$0.80$, so $61$ of the $66$ \textit{Need Practice} recordings fall inside the range occupied by \textit{Great} ones (Figure~\ref{fig:calibration}a).


Given this overlap, the dashboard assigns three labels based on a dual threshold ($\leq 0.02$ and $> 0.20$) calibrated against SLP consensus.
Since the calibration set is $41\%$ non-disordered by design, this distribution reflects the set's construction (Figure~\ref{fig:calibration}b).
Measured against the same SLP consensus throughout, the \textit{Great} zone matches consensus on $95.5\%$ of its recordings and the \textit{Need Practice} zone on $83.7\%$.
The \textit{Marginal} label denotes the interval in which the rule withholds an automated \textit{Great}/\textit{Need Practice} decision; it should not be read as the system identifying a clinically intermediate production, even though the SLP scale uses the same term.
Within this abstention band, $28$ of the $43$ recordings ($65.1\%$) were rated \textit{Need Practice} by the SLPs.

Because the rule abstains rather than classifying every production, its behavior is best described as selective prediction.
Restricted to the $120$ recordings whose SLP consensus is \textit{Great} or \textit{Need Practice}, forcing a binary verdict at the best single cut point agrees with consensus on $88.3\%$; the dual threshold instead labels $70.0\%$ of them and agrees on $92.9\%$ of those, withholding judgment on the rest.\footnote{Agreement with the SLP1--SLP2 consensus is in-sample, because those ratings set the thresholds. 
A third licensed SLP, who did not participate in calibration, independently rated the same 130 recordings; against this rater, the fixed thresholds yielded 70.8\% coverage and 92.0\% selective accuracy~(vs. 70.0\% and 92.9\% against the calibration consensus), with 86.4\% agreement inside the automated \textit{Great} zone~(95.5\%).
This provides an independent-rater check on the same recordings, not validation on unseen recordings or speakers, and each figure should be interpreted against its own rating reference.}
Recordings remain available in every label category.
The agreement between the SLPs describes their aligned ratings, not the accuracy of the automated labels, and the thresholds were calibrated on STAR recordings: this analysis does not establish their performance on the child--caregiver study recordings or the reliability of the separate target-registration mechanism.


%% file: sections/5-evaluation.tex
\section{Child--Caregiver User Study}
\label{sec:user-study}
We examined \sysname{} by following three research questions:
\begin{enumerate}
    \item \textbf{RQ1}. How did children respond to \sysname{}'s story-framed practice modes, and what comprehension or formulation challenges accompanied their productions?
    \item \textbf{RQ2}. How did child--caregiver dyads interpret listen-back and target-registration feedback and manage further attempts?
    \item \textbf{RQ3}. What caregiver involvement occurred during practice, and what roles and conditions did families anticipate for home use? 
\end{enumerate}

The study was reviewed and approved by our institution's IRB.

\subsection{Study Descriptions}

\subsubsection{Participants and Configurations}

\begin{table*}[t]
\centering
\setlength{\tabcolsep}{4pt}
\begin{tabular}{c cc cc c l c c c}
\hline
& \multicolumn{2}{c}{\textbf{Caregiver}}
& \multicolumn{2}{c}{\textbf{Child}}
& & & & & \\
\cline{2-3}\cline{4-5}
\textbf{Dyad ID}
& \textbf{Gender} & \textbf{Age}
& \textbf{Gender} & \textbf{Age}
& \textbf{Group}
& \textbf{Language at Home}
& \textbf{Sound(s)}
& \textbf{\# Syll.}
& \textbf{Mode} \\
\hline

1  & F & 35--44 & M & 7
    & No therapy
    & English, Mandarin
    & initial /r/ & 2 & Conversational \\

2  & F & 45--54 & M & 11
    & No therapy
    & Mandarin
    & initial /l/ & 2 & Word/Sentence \\

3  & M & 25--34 & M & 11
    & SSD/history
    & Mandarin
    & initial /s/ & 3 & Both \\

4  & F & 35--44 & M & 11
    & No therapy
    & English, Mandarin
    & initial /r/ & 3 & Both \\

5  & F & 35--44 & F & 6
    & No therapy
    & English
    & initial /l/, /r/ & 2 & Both \\

6  & F & 35--44 & F & 8.5
    & No therapy
    & Mandarin
    & initial /z/ & 2 & Conversational \\

7  & F & 35--44 & M & 6
    & SSD/history
    & Mandarin
    & initial /s/ & 2 & Conversational \\

8  & F & 45--54 & F & 8
    & SSD/history
    & Mandarin
    & initial /s/, /\textesh/ & 2 & Word/Sentence \\

9  & F & 35--44 & M & 10
    & SSD/history
    & English
    & initial /s/ & 2 & Word/Sentence \\

10 & F & 45--54 & M & 7
    & SSD/history
    & English
    & initial /r/ & 2 & Conversational \\
\hline
\end{tabular}

\caption{Demographic information and study configurations for the
child--caregiver dyads. ``No therapy'' denotes no reported prior speech-therapy experience; ``SSD/history'' denotes a reported SSD diagnosis or prior speech-therapy experience. Sound(s), syllable complexity, and mode indicate the study materials prepared for each dyad.}
\label{tab:evaluationparticipant}
\end{table*}

We recruited 10 child--caregiver dyads through purposeful sampling via word of mouth~\cite {palinkas_purposeful_2015}~(Table~\ref{tab:evaluationparticipant}).
We refer to caregivers as P1–P10 and children as C1–C10 by dyad number.
All dyads reported they were interested in speech-sound practice: five children had no prior speech-therapy experience (\notherapykid{1}, \notherapykid{2}, \notherapykid{4}, \notherapykid{5}, \notherapykid{6}) while five comprised the SSD/therapy-history group, having either a reported SSD or prior speech-therapy experience (\yestherapykid{3}, \yestherapykid{7}, \yestherapykid{8}, \yestherapykid{9}, \yestherapykid{10}).
We use these groupings descriptively, not as clinically equivalent or experimentally controlled cohorts.
Seven of ten caregivers reported Mandarin in the home, five of them exclusively, and three reported English only. 
Several configured targets---initial /l/, /z/, /r/---name contrasts on which second-language phonological transfer and a speech sound difficulty predict overlapping production patterns.
We did not collect language-history measures that would separate them, and we do not treat any observed non-target production as evidence of a disorder.
Because participants' treating SLPs were not available to the study, the research team prepared two story materials per dyad from caregiver-reported target sounds or blends, specific words, and preferred practice levels.
We compensated participants with \$40 USD for their participation.

\subsubsection{Procedure}
Each dyad attended one approximately 60-minute visit at our university.
Caregivers provided written informed consent on behalf of their child; children provided verbal assent.
Caregivers were asked to let the child lead, intervening when the child had difficulty understanding or progressing.
Each dyad completed at least one story-based activity and could continue with a second, using Word/Sentence Mode, Conversational Mode, or both by preference; modes were not assigned as comparison conditions.
A brief child debrief and a semi-structured caregiver interview then addressed the appropriateness of the content, experiences with the narrative and feedback, caregiver involvement, and how the activity might fit home practice, followed by a questionnaire based on the Technology Acceptance Model~\cite{venkatesh_technology_2008}.
The screen and audio were recorded throughout the session. 
\footnote{Third-party APIs used to process recordings are covered under our institution's enterprise subscriptions to these services, which carry HIPAA/FERPA-compliant data processing agreements; under these agreements, recordings and derived data are not used by the providers to train their models.}
As a single-visit laboratory study, it examined interaction with the system and anticipated home use rather than longitudinal adherence, clinical outcomes, or actual use in the home.

\subsubsection{Analysis Methods}
\label{sec:child-analysis-methods}

\begin{table*}[t]
\centering
\caption{Units for story-practice counts across 10 dyads in Section~\ref{sec:findings}. Sidebar practice is excluded.}
\label{tab:units}
\begin{tabular}{@{}p{2.6cm}p{7.2cm}p{2.6cm}@{}}
\toprule
\textbf{Unit} & \textbf{Definition} & \textbf{N} \\
\midrule

Offered opportunity &
A distinct story task requesting a target-bearing response.
Includes bypassed tasks; retries and re-presentations remain within the same opportunity.
Denominator for caregiver support and retry occurrence.&
127 \\[2pt]

Response-formulation task &
A subset of offered opportunity requiring sentence production (Word/Sentence) or an open/framed response (Conversational), including framed choices.
Excludes word slots, item tasks, and unframed choices.
Denominator for response realization. &
61 (21 W/S; 40 Conv.) \\[2pt]

Retry episode &
A task presentation with an initial child recording and at least one additional recording to the same task, including a changed choice.
No opportunity contained more than one coded episode.
Denominator for additional recordings per episode. &
53 \\

\bottomrule
\end{tabular}
\end{table*}

We employed mixed methods to characterize how children and caregivers engaged with \sysname{}. 
First, we analyzed the screen and audio recordings together with the available interaction records to quantify practice behavior~(Table~\ref{tab:units}), reporting descriptive statistics only: given the small, heterogeneous sample, we made no inferential comparisons between children with and without therapy experience.
Second, a Technology Acceptance Model~\cite{venkatesh_technology_2008}-based 7-point post-questionnaire completed by caregivers, summarized at the item level; nine questionnaire records were available for analysis because \yestherapy{10}'s survey response was not retained.
Third, the debriefing interviews were analyzed with thematic analysis~\cite{braun_reflecting_2019, braun_using_2006}.
Transcripts were segmented into meaning units, the first coder iteratively open-coded them into a codebook addressing the research questions, a second coder independently applied the codebook without seeing the first coder's labels, and disagreements were reconciled into the final codes.
We examined patterns across the behavioral measures, interview themes, and survey responses to identify points of convergence and tension.
Survey responses provide context for how caregivers received the system; the findings below are grounded primarily on observed interactions and debriefing accounts.

\subsection{Findings}
\label{sec:findings}
The behavioral counts below are descriptive summaries of codes applied to story practice, referenced to the 127 offered opportunities in Table~\ref{tab:units}, and three limits apply throughout. 
First, they describe elicitation and support, not articulatory accuracy: no count records whether a production was correct. 
Second, modes were self-selected and dyads differed in age, target, and language background, so no count establishes a mode effect or a between-participant difference (Section~\ref{sec:child-analysis-methods}). 
Third, retry-episode counts include changes of response choice and describe re-recording, not pronunciation errors or independently assessed improvement.

\subsubsection{\textbf{RQ1}. How did children respond to \sysname{}'s story-framed practice modes, and what comprehension or formulation challenges accompanied their productions?}
\label{sec:study-production}
The story and configured interaction mode shaped children's speaking opportunities without fully determining the demands they encountered.
Story framing could make speaking consequential while adding comprehension demands, and nominal linguistic levels still varied in vocabulary, reading, and formulation requirements.

\paragraph{Story-framed practice created reason to speak, but also more to understand}
Six out of 10 children identified a character, theme, challenge, or story event as a favorite element of this practice.
For some, the valued story elements were also occasions to speak: \yestherapykid{7} selected the battle scene as a favorite because ``\textit{I have time to talk with other people}'', while \yestherapykid{9} valued that the story ``\textit{keeps on giving like challenges}''.
Speaking was not presented only as repetition of a target word; it was one of the actions through which the child met a character, made a choice, or completed a challenge.

The same event, however, could increase both the reason to speak and the amount the children had to understand.
\yestherapykid{7} also found the sudden transition in battles hard to follow initially, ``\textit{I had no idea what they're doing in the story}''.
\yestherapy{8} likewise judged the target-word practice itself useful while finding the surrounding story ``\textit{a little bit too difficult for her}'', preferring shorter and simpler language around those targets.
A phoneme-dense narrative could become a linguistic burden even when the child found it engaging: narrative appeal and understanding a particular speaking task were related but distinct.
The goal is therefore not simply to make speech practice \textit{more story-like}, but to preserve that narrative reason to speak without making story comprehension a prerequisite for speaking.

\paragraph{The same level asked caregivers for different kinds of work}

\begin{figure}
    \centering
    \includegraphics[width=\linewidth]{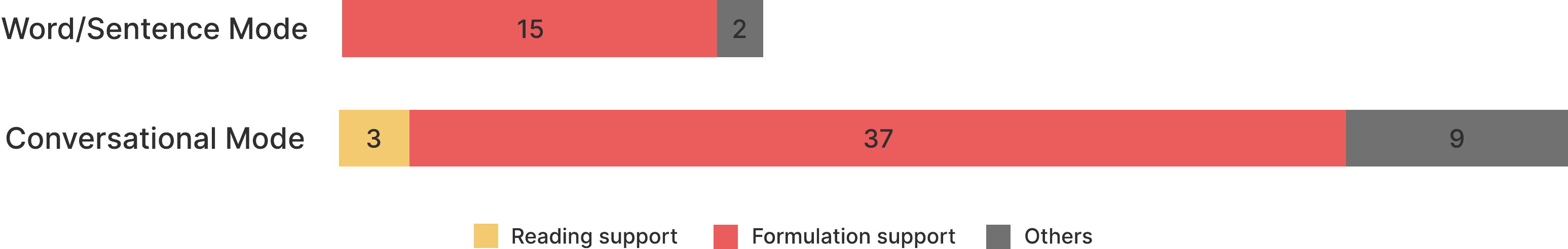}
    \caption{Caregiver support applications by mode across the 127 offered story opportunities (64 coded support acts, 66 nonexclusive category applications). Reading support accounted for 15 of the 17 applications in Word/Sentence Mode; formulation support for 37 of the 49 in Conversational Mode. ``Other'' groups included question clarification (9), vocabulary explanation (1), and plot clarification (1). Counts describe coded acts, not the proportion of opportunities completed independently, and modes were self-selected.}
    \label{fig:support}
\end{figure}

The modes constrained responses differently, but the resulting utterances also depended on support during the task.
Of the 61 response-formulation tasks (Table~\ref{tab:units}), 19 of 21 sentence-production tasks in Word/Sentence Mode ended in a complete target-bearing sentence, including sentences modeled by an adult, whereas 17 of 40 open or framed tasks in Conversational Mode ended in connected, child-contributed content.
The support caregivers supplied differed in kind rather than degree (Figure~\ref{fig:support}): nearly all support in Word/Sentence Mode was reading support, whereas most support in Conversational Mode was formulation support, with question clarification less frequent and vocabulary or plot explanation rare.
The gap between a configured mode and the realized task was sometimes visible in the caregiver's own correction: after the child produced a full sentence, \notherapy{5} clarified that this particular turn wanted only the single word.

Some dyads found the task difficult even in a mode they had chosen as appropriate.
\yestherapy{8} selected Word/Sentence Mode expecting ``\textit{just one short sentence or a word}'', like the short constructions of current therapy such as ``\textit{the boy is playing}''.
Calling both the word-filling and sentence-production tasks ``sentence-level'' practice therefore concealed differences in utterance length, lexical load, and the reading and production support required.
Other mismatches came from vocabulary rather than the target sound: \yestherapy{3} judged the child's production ``\textit{not too bad}'' but found words such as \textit{Starbucks} and \textit{strolled} difficult enough to propose a brief vocabulary preview before practice to reduce the burden of understanding words and producing sounds together.
Three caregivers, including \yestherapy{7} and \notherapy{6}, wanted vocabulary grounded in what the child encountered at school and at home.
Configuring the target sound and response format did not fully specify the vocabulary, reading, and formulation demands children encountered.


\subsubsection{\textbf{RQ2}. How did child--caregiver dyads interpret listen-back and target-registration feedback and manage further attempts?}
\label{sec:study-feedback}
Feedback gave dyads evidence with which to inspect a production and opportunities to try again.
Caregivers received feedback as a resource for attending to speech, agreeing that it made them talk with their child about how to speak (e.g., volume, tongue position) more than usual~(M=6.00, SD=1.00).
That reception set up a coordination task rather than a settled verdict: the signal shaped a production's interpretation without settling it, leaving the dyad to judge when to stop trying.

\paragraph{Feedback made production inspectable without resolving correctness}
Listen-back first made the child's own production available for inspection.
\notherapy{1} explained hearing oneself could reveal, ``\textit{There's some part that, oh, I didn't realize I was not able to pronounce it well}''.
\yestherapy{3} recalled the child heard a prior recording and saying, ``\textit{Oh, I said that?}''
Exposing a production therefore created material for reflection before either the system or the caregiver had resolved what it meant.
The target-registration feedback was more directive, pointing at a specific production to trigger another attempt, 
\notherapy{5} valued this specificity: ``\textit{if you want to practice a specific pronunciation or any sound, you have the option here, and you can emphasize that sound by going through the steps.}''

That same specificity made target-registration easier to read as a verdict than it was designed to be.
\yestherapy{9} described the red and green markings as showing ``\textit{what he got right and what he missed}'', so that ``\textit{he knows how to fix it}''---a broader judgment about pronunciation, not just a pointer toward one more attempt.
One family wanted the more explicit judgment: \notherapy{4} proposed labels such as \textit{excellent job} or \textit{well done} or \textit{need improvement}, while \notherapykid{4} wanted the interface to name the particular word that needed attention; otherwise, the child reasoned, users ``\textit{probably wouldn't know how to improve}.''
We treat this as a particular preference rather than a common requirement.
In the opposite direction, recognition failures could weaken the authority of the signal:
\notherapykid{6} described a difficult point as one ``\textit{where the computer messes up and I have to restate it and restate it}''---the repair pattern Beneteau~\textit{et al.}~\cite{beneteau_communication_2019} documented in families' everyday agent use, here recurring inside a practice task.
Families thus noticed a prior production, read a signal as a broader judgment than intended, wanted that judgment made explicit, or attributed a mismatch to recognition rather than to the child.
Keeping phonemic labels out of the practice interface did not, by itself, prevent families from reading its remaining signals as evaluations.

\paragraph{Retry was productive up to a boundary families set differently}
Of the 127 offered story opportunities (Table~\ref{tab:units}), 53~(41.7\%) included a retry episode.
These 53 episodes occurred in all ten dyads, with three to eight episodes per dyad, and contained 98 additional child recordings (M=1.85 per episode, SD=1.22; median=1, range=1--7).
Two caregivers described benefits of the retry process.
\yestherapy{10} recalled the system saying ``\textit{try again}'', the following production sounding ``\textit{a little bit better}'' to her, and ``\textit{[the system] actually picked up on him doing better}'': the caregiver heard a change, and the interface responded differently.

Families disagreed, however, on how far the loop should continue.
\notherapy{1} preferred a strong progression contingency, ``\textit{not letting him proceed to the next step until he corrected it}'', whereas \yestherapy{8} worried that repeated retry requests after an unsuccessful production could undermine the child's confidence, contrasting them with the child's therapist, who stayed encouraging---``\textit{good, good job}''---even when the production was not yet successful.
We do not infer that prior therapy experience causes these preferences; their contrast exposes a calibration problem.
The disagreement was not whether repetition mattered but when another attempt remained productive: it can remain useful practice or become another encounter with failure, and neither a universal retry count nor a target-registration signal can locate that boundary by itself.
The system can surface a possible mismatch, replay the recording, invite another attempt, and retain the production; the child can accept another attempt or move on; and caregivers can observe willingness, fatigue, and frustration and help regulate persistence without becoming pronunciation judges, since unresolved productions remain available for later professional review.

\subsubsection{\textbf{RQ3}. What caregiver involvement occurred during practice, and what roles and conditions did families anticipate for home use?}
\label{sec:study-support}
Caregivers rated that the app could fit their daily routine~(M=5.56, SD=1.13), reducing setup or monitoring effort~(M=5.89, SD=0.78), and being less burdensome than their current materials (M=6.22, SD=0.67); these ratings describe reception after one supported session.
The debriefs identified what would have to hold at home.


\paragraph{Anticipated independence depended on preparation, initiation, and follow-up.}
Families described different degrees of child autonomy, but even their most independent scenarios were conditional.
\yestherapy{3} believed the child could ``\textit{do the whole thing without other people helping}'' but tied that to first providing vocabulary explanations and visual aids; \notherapy{1} imagined the child practicing while she cooked and returning to a session report: ``\textit{I don't have to spend the whole 30 minutes with him. I can do something by myself and then come back}''.
Independence meant replacing continuous co-presence with preparation or later review, not removing the caregiver from the practice cycle.
Initiation, in particular, was expected to remain adult work: \yestherapy{9} questioned whether the child would independently choose optional word practice, though the child responded, ``\textit{I probably would have done the word that I want}''---an interest in choosing words, distinct from initiating and organizing practice.
Routines were already constrained: \yestherapy{9} practiced around several children and after-school activities, and \yestherapy{10} envisioned a bounded slot before school in which the child would participate ``\textit{if I had to have him do it}''.
Being able to operate the interface therefore did not imply independently remembering, initiating, or sustaining practice.

Families also looked beyond the immediate dyad for direction.
\notherapy{5} wanted a specialist to recommend ``\textit{how long, what should we do}'' and the ``\textit{step-by-step}'' order of practice, and \yestherapy{9} wanted brief weekly comments or tips: requests for intermittent direction to make recorded practice actionable, rather than clinician presence at every attempt.
Access to that professional support was itself conditional: \yestherapy{10} described out-of-pocket costs that currently prevented in-person services and imagined using \sysname{} as practice support while such services were unavailable.


Overall, families envisioned AI mediation as redistributing work across time: children could carry out portions of the activity, caregivers could prepare, initiate, schedule, and later review, and professionals could provide direction and follow-up.
Its anticipated value was a reduced need for continuous caregiver co-presence or correction, not removal of the coordination needed to make practice happen and decide what comes next.



\paragraph{Immediate acceptance came with conditions for continued use}
The acceptance ratings above describe reception after a single supported session and are reported only as context.
For \yestherapy{9}, novelty was both the draw and the limit: once the activity became familiar, she doubted the child would remain interested or begin without prompting, and asked for shorter stories and a clear time limit; asked whether he would practice every day, \yestherapykid{9} answered, ``\textit{If you want me to}''.
Two children named the generated voice as a reason it might not last---a ``\textit{robot}'' (\yestherapykid{9}), ``\textit{more flat}'' (\notherapykid{2})---consistent with children's documented sensitivity to synthetic voice~\cite{lovato_siri_2015}.
Caregivers also bounded the obvious remedy: 
\yestherapy{7} emphasized that this was \textit{``not a game app''}, and \notherapy{6} supported only a small amount of added gaming rather than letting games displace practice. 
Continued use therefore turned less on making the activity more engaging than on whether anyone would start it---which returns the question to the caregiver.

%% file: sections/6-discussion.tex
\section{Discussion}
\label{sec:discussion}
\sysname{} connects story-based elicitation, support during an attempt, and contextualized
records within one practice system.
Building and studying it across three stages surfaced three tensions that any system placing
speech practice in a story, a signal, and a child and caregiver will meet: response constraint and
comprehension load vary independently; a signal's evaluative weight follows its immediacy
rather than its intended scope; and reducing adult correction raises the demand for adult
coordination.

\subsection{Shared and Revisable Configuration}
\label{sec:discussion-roles}

The two studies, the expert walkthrough and the child--parent study, revealed the need for configuration at different points. 
First, the knowledge needed to set both axes is distributed.
An SLP can define target sounds, production contexts, and acceptable ranges of complexity; a caregiver can supply familiar vocabulary and adjust length or setting within those bounds; children can make choices they can understand, such as which activity to attempt or when to stop—an affordance \sysname{} does not currently provide beyond destination choice and retry, to facilitate intrinsic motivation and self-regulation~\cite{10.1145/3544548.3580935, 10.1145/3773077.3806105}. 
These contributions are not interchangeable, and a shared configuration space should distinguish expert-set constraints from family preferences rather than exposing every parameter to everyone. 
Precision should also not require assembly: presets and generated word sets that remain inspectable and selectively editable let an SLP refine one constraint without rebuilding an activity, and let a caregiver contribute words without navigating a phonetic interface.

Second, configuration should remain revisable during use, because the mismatches that mattered surfaced only in the moment of practice. 
Caregivers' explanations, reformulations, and decisions to stop were observable evidence of task demand that no initial setting captured. 
A system could let a caregiver mark an unfamiliar word, an over-long sentence, or a decision to stop after repeated attempts, and attach that mark to the relevant recording. 
This asks caregivers to report what they did, not to diagnose why a production failed, and it makes repeated mismatches available to the next configuration decision instead of leaving them to be resolved each session afresh.

\subsection{Calibrating AI’s Authority to Scope-Bounded Feedback}
\label{sec:discussion-ai-authority}

AI-mediated speech practice requires separating three functions: generating an activity, registering a response, and evaluating speech production. 
In \sysname{}, these are connected in the interaction but do not provide interchangeable evidence: generated dialogue may contain the requested targets while imposing excessive vocabulary or formulation demands; a target word registered by ASR does not establish articulatory accuracy; and ASR artifacts can dominate phonemic scores precisely where actual error is small (Section~\ref{sec:scoring}).
Competence in one function should not confer authority over another: generation can vary content within specified constraints, recognition can report what was registered, and scoring can offer provisional information for professional review.

Separating audiences is necessary but insufficient to maintain these boundaries.
The expert walkthrough showed compact summaries could support selective review without settling production quality (Section~\ref{expert:review}); the child--caregiver study showed target-registration feedback was read either as a pronunciation judgment or as a recognition failure (Section~\ref{sec:study-feedback}).
Keeping quality labels off the practice screen therefore did not remove evaluative feedback from the child's experience; it left one immediate, high-frequency signal in place---a feedback pattern associated with momentary performance gains but weaker retention~\cite{maas2008principles}. 
Placing an output by audience is therefore a weaker control than regulating what the child-facing interaction signals at all.
The obvious remedy inherits the problem: surfacing transcripts and mismatch reasons would make the pipeline's inconsistencies visible, but explanations increase reliance on incorrect outputs as well as correct ones~\cite{10.1145/3706598.3714020, 10.1145/3581641.3584066}.



Accordingly, systems should calibrate the consequences of automated outputs, not only how confidence is displayed.
An unregistered target can justify offering another attempt without justifying an indefinite progression gate, a change in practice level, or a claim that the child needs more corrective practice.
Bounded retries, options to pause or continue, and preservation of unresolved recordings offer a way to maintain practice flow without treating progression as evidence of correct production.
For professionals, uncertainty should likewise be attributed to the system's inference rather than to the child's ability.
The calibrated \textit{Marginal} category, for example, denotes a range where automated classification is uncertain, not an established level of borderline production quality.
Making that distinction explicit and retaining access to the original audio are important complements to a compact label.
Better models are complementary: sound-specific classifiers such as Benway~\textit{et al.}'s rhoticity detector~\cite{Benway2023ProspectiveVO} would strengthen the available evidence. 
But classification accuracy does not determine whether an output should invite a retry, gate progression, or change the task, nor when the target should leave home practice altogether.
The formative warning that unsupported repetition can entrench incorrect movements (Section~\ref{formative:caregiver}) names a risk the score cannot weigh (Section~\ref{sec:scoring}). 
That judgment is the clinician's, and the records already hold its prompts---sustained \textit{Need Practice} labels on one target, opportunities repeatedly advanced without a match---making suspension of that target one configuration outcome alongside adjusting or continuing; the child-facing activity does not change.

Finally, scope-bounded feedback must also apply to the records that preserve automated outputs.
Because recordings and labels can outlast the interaction, and our thresholds were calibrated on a population that did not match much of our study sample (Section~\ref{sec:limitations}), deployment should preserve calibration provenance and original audio, require clinician sign-off before automated labels are used as assessment evidence, and specify access, retention, deletion, and consent for any reuse of recordings.

\subsection{Support for Coordination, Not Only Correction}
\label{sec:discussion-coordination}
Families described independence as a redistribution rather than a removal, the pattern joint-media-engagement research describes for co-present parents~\cite{hiniker_lets_2018, garg_he_2020}.
\sysname{} absorbed correction at the moment of production---it supplied the model, the cue, and the request for another attempt.
What it did not absorb was deciding when practice happens, whether another attempt is worth asking for, and what should come next.
That residual work has to land somewhere, and each candidate destination in our studies was already constrained.
Clinicians were the natural destination for decisions requiring expertise, yet the expert walkthrough showed that routine review of recorded practice had to fit within limited clinical attention (Section~\ref{expert:review}). 
Caregivers were the destination for scheduling and initiation, and \yestherapy{9} described fitting practice around several children and after-school activities while \yestherapy{10} pictured a single bounded slot before school.
Children were the destination for initiation itself, and neither \yestherapy{9}'s doubt nor \yestherapykid{9}'s answer to whether he would practice daily---``\textit{If you want me to}''---suggested it would originate there.
Reducing correction while relocating coordination therefore does not obviously reduce total effort; it changes who is asked for what, and when.

This reframes what such systems should automate.
Coordination that requires no judgment---remembering an assignment, surfacing the next step, bounding a session---is a better target for automation than pronunciation judgment, which our scoring analysis indicates the pipeline cannot deliver reliably in any case.
Coordination that does require judgment should be batched into the smallest action the responsible person can afford: \expert{1} proposed a weekly summary email rather than per-child inspection, and \yestherapy{9} asked for weekly ``\textit{fast comments}'' rather than continuous contact.
Both describe intermittent, low-cost re-entry rather than sustained supervision.

The same logic bounds the access claim these systems invite.
Families' accounts showed that professional availability and affordability could constrain what support accompanied practice (Section~\ref{sec:study-support}), while \expert{6} raised families' differing access to devices and experience with technology.
An available record is not the same as an available reviewer, and access to an application is not access to professional services.

%% file: sections/7-limitation.tex
\section{Limitations and Future Work}
\label{sec:limitations}


This work has several limitations.
The user study included ten child--caregiver dyads: five children with a reported SSD or prior speech-therapy experience and five without prior speech-therapy experience.
Recruiting families who met the inclusion criteria and were willing and able to participate constrained the size of the SSD/therapy-history group.
These descriptive groupings do not establish comparable clinical profiles, and the small, heterogeneous sample limits how broadly the observed interaction patterns can be extended to children with different speech and support needs.

We were also unable to involve participants' treating SLPs in configuring the study materials, and some families had paused therapy.
The research team therefore configured the stories using caregiver-reported targets and difficulty preferences rather than clinician-confirmed practice plans.
Although this allowed us to incorporate families' stated practice interests, the selected targets and linguistic levels may not have fully matched each child's current practice needs.
Observed difficulty may consequently reflect the fit of the configured materials as well as the interaction design.
Future studies should involve families and treating SLPs in establishing and revising these practice parameters.

Each dyad participated in a single researcher-supported laboratory visit, rather than routine home use.
Positive reception and observed caregiver involvement therefore do not establish sustained engagement, independent home use, or reduced caregiver burden.
Moreover, the intended coordination cycle---configure, practice, review, adjust---was not evaluated as an ongoing workflow.
A longitudinal deployment with children diagnosed with SSD and their treating SLPs is needed to examine whether engagement is sustained, how caregiver effort changes, and whether SLPs integrate dashboard records into ongoing clinical reasoning and subsequent configuration.
Our behavioral measures describe production opportunities, attempts, and target-registration rather than independently assessed improvements in articulatory accuracy.
Comparative studies with independently assessed speech outcomes would be needed to evaluate relative effectiveness and generalization gains.

Our sample was linguistically heterogeneous in a way we could not analyze. 
Seven of ten households reported Mandarin at home, and five reported Mandarin only, while the scoring thresholds were calibrated on different language bases. 
We did not collect age of English exposure, relative language dominance, or caregiver-rated accentedness, and we did not obtain independent clinical judgments of the study recordings. 
Observed registration failures and observed difficulty may therefore reflect the fit of the configured materials, second-language phonological patterns, recognition degradation on accented child speech, or a speech sound difficulty, in proportions we cannot estimate. 
Future studies should either recruit within a single language background or collect language-history measures sufficient to model it, and should validate scoring thresholds on speech from the population the system is deployed to.

The study also covered a limited set of target sounds and narrative templates.
Findings concerning narrative appeal and comprehension demands may therefore depend partly on the particular stories and vocabulary used.
Future work could expand this coverage and explore difficulty adjustment informed by caregiver input and clinician-reviewed practice histories, rather than target-registration success alone.
Extension to other communication disorders, such as developmental language disorder, remains a longer-term design opportunity requiring separate evaluation.


%% file: sections/8-conclusion.tex
\section{Conclusion}

We presented \sysname{}, an AI-mediated role-playing system that combines target-bearing story interactions, practice cues, listen-back, and contextualized recordings for clinician review.
Clinician involvement shaped its design requirements, while a separate scoring analysis examined the limits of automated dashboard labels.
In a study with ten child--caregiver dyads, narrative participation required varying language support, target-registration signals acquired evaluative meanings, and caregivers anticipated redistributing support across preparation, practice, and follow-up.
These findings inform more interpretable and support-aware speech-practice systems, while sustained home use and the full clinician--home workflow remain for future evaluation.